\documentclass[conference]{IEEEtran}
\IEEEoverridecommandlockouts
\usepackage{cite}
\usepackage{amsmath,amssymb,amsfonts}
\usepackage{graphicx}
\usepackage{textcomp}
\usepackage{xcolor}
\usepackage{enumitem}
\usepackage{algorithm}
\usepackage{flushend}
\usepackage{algpseudocode}
\usepackage{listings}
\usepackage{minted}
\usepackage{flushend}
\usepackage{todonotes}
\usepackage{subfloat}
\usepackage{multirow}
\usepackage{multicol}
\usepackage[font=small,labelfont=bf]{caption}
\usepackage{subcaption}
\usepackage{subcaption}
\usepackage{setspace}
\usepackage{xcolor}
\usepackage{subfig}

\AtBeginDocument{%
  \providecommand\BibTeX{{%
    Bib\TeX}}}
\def\BibTeX{{\rm B\kern-.05em{\sc i\kern-.025em b}\kern-.08em
    T\kern-.1667em\lower.7ex\hbox{E}\kern-.125emX}}
\begin{document}

\title{Cross-Layer Energy Analysis of Multimodal Training on Grace Hopper Superchips\\
}
\author{
Mahmoud Ahmed\textsuperscript{1},
\IEEEauthorblockN{Sameh Abdulah\textsuperscript{1}, 
Olatunji Ruwase\textsuperscript{2},
Sam Ade Jacobs\textsuperscript{3}, \\
Mathis Bode\textsuperscript{4},
Mohamed Elhoseiny\textsuperscript{1},
David E. Keyes\textsuperscript{1}
\vspace{4mm}}

\IEEEauthorblockA{\textsuperscript{1}Computer, Electrical and Mathematical Sciences and Engineering (CEMSE) Division, KAUST, Thuwal, Saudi Arabia}
\IEEEauthorblockA{\textsuperscript{2}Snowflake, Bellevue, WA, USA}
\IEEEauthorblockA{\textsuperscript{3}Microsoft Inc.}
\IEEEauthorblockA{\textsuperscript{4}J\"ulich Supercomputing Centre, Forschungszentrum J\"ulich GmbH, Germany}
}

\maketitle

\begin{abstract}
Multimodal deep learning models enable joint learning across heterogeneous data sources, including text, images, and video, but their rapid scaling introduces significant memory and communication bottlenecks. As model sizes and sequence lengths increase, training performance becomes increasingly impacted by data movement rather than computation. Frameworks such as DeepSpeed mitigate these challenges through CPU offloading, activation checkpointing, and communication optimizations. However, these techniques introduce additional system activity, which may affect energy efficiency. Meanwhile, tightly integrated heterogeneous architectures, such as the NVIDIA Grace Hopper (GH200) superchip, provide high-bandwidth CPU-GPU interconnects and unified memory, thereby reducing data transfer overhead. In this work, we present a cross-layer analysis of energy and performance trade-offs in multimodal training on GH200 systems, explicitly characterizing the interactions between application, runtime, and hardware layers. Leveraging high-bandwidth CPU–GPU interconnects, our results show that energy efficiency is primarily governed by data movement and overlap rather than raw compute utilization, and that configurations optimized for runtime are not necessarily optimal for energy. Based on these findings, we distill a set of actionable guidelines for practitioners that demonstrate how to balance offloading strategies, sequence parallelism, and hardware-aware scheduling to achieve energy-efficient training. Our results demonstrate that leveraging high-bandwidth CPU-GPU interconnects enables offloading strategies and sequence parallelism, achieving a strong balance among energy efficiency, runtime performance, and computational throughput, providing practical guidelines for efficient multimodal training on modern heterogeneous systems.
\end{abstract}


\section{Introduction}

Large multimodal models that combine vision encoders, temporal processing modules, and large language models (LLMs) have grown rapidly in scale and complexity, driven by the need for richer contextual understanding in diverse applications~\cite{wang2023large,zhang2024mm,carolan2024review}. Recent progress in video-based multimodal LLMs has accelerated this growth, as video inputs, represented as frame sequences, require extensive encoding, storage, and processing over time~\cite{lin2023videollava,damonlpsg2023videollama}. This leads to higher token counts and larger intermediate feature sets, increasing memory demands and inter-device communication. 
While advances in accelerator architectures and parallel training frameworks have improved computational throughput, model efficiency is now increasingly limited by data movement across memory hierarchies and interconnect systems~\cite{rajbhandari2020zero}. 
Although prior work has focused on scaling LLMs and optimizing distributed training performance, the energy behavior of multimodal workloads on tightly coupled CPU–GPU systems remains poorly understood. Existing studies emphasize throughput and memory efficiency~\cite{rajbhandari2020zero,smith2022using,hielscher2025optimizing}, leaving the impact of system architecture and runtime strategies on energy largely unexplored.

Energy efficiency is now a first-class design constraint in contemporary supercomputing environments, where large-scale AI workloads directly impact system sustainability, operational cost, and deployment scalability~\cite{cao2023reducing,dahule2024analyzing,wang2025environmental,alomairy2025sustainably}. As AI-driven applications increasingly dominate HPC usage, their computational intensity and extended training cycles are expected to increase overall energy demand significantly. At the same time, modern platforms are becoming increasingly heterogeneous, integrating CPUs, GPUs, deep memory hierarchies, and high-speed interconnects. This evolution fundamentally complicates system behavior, as energy consumption is no longer dictated solely by compute units but is increasingly influenced by memory accesses and data movement across system components. 
In this context, modern hardware innovations, such as the NVIDIA Grace Hopper (GH200) superchip and its successors, introduce a new class of tightly coupled heterogeneous architectures that integrate CPU and GPU resources via high-bandwidth interconnects and unified memory, thereby alleviating data movement bottlenecks. While this design improves data locality and programmability, it also introduces more intricate energy dynamics driven by interactions between compute and memory subsystems. By tightly coupling CPU and GPU through a high-bandwidth, cache-coherent interconnect and unified address space, GH200 enables efficient cross-device data sharing. This integration introduces additional energy overheads due to coherence, cross-device communication, and data placement across HBM and LPDDR, making power behavior dependent on coupled interactions among compute, memory, and interconnects~\cite {muralidhar2022energy}.

Modern deep learning frameworks, such as DeepSpeed~\cite{rasley2020deepspeed}, further leverage these architectural features to enable memory- and communication-aware optimizations for large-scale training. However, these features significantly reshape execution patterns, making the relationship between performance and energy increasingly complex and less intuitive. DeepSpeed introduces advanced optimization techniques, including ZeRO-based optimizer offloading, activation checkpointing~\cite{rajbhandari2020zero,rajbhandari2021zero}, and sequence parallelism~\cite{jacobs2023deepspeed}. While these mechanisms improve scalability, they fundamentally reshape execution by redistributing computation and data movement across CPU, GPU, and memory hierarchies. For example, offloading reduces GPU memory pressure but increases CPU-GPU communication, while activation checkpointing trades memory usage for recomputation ~\cite{chen2016training,rajbhandari2020zero,narayanan2021efficient}. These transformations introduce nontrivial trade-offs between computation, communication, and memory access, making energy consumption difficult to predict. Traditional performance metrics, such as throughput (FLOP/s) and time-to-solution, are insufficient to capture these trade-offs. Configurations that maximize computational efficiency may incur higher communication overhead or memory traffic, thereby increasing energy. Despite the growing importance of energy efficiency in large-scale AI systems, there is a limited understanding of how runtime strategies interact with heterogeneous architectures across different system layers. In particular, the relative contributions of computation, memory access, and interconnect traffic remain insufficiently characterized for multimodal workloads.

Building on these challenges, this work provides a cross-layer analysis of energy and performance trade-offs in multimodal training on GH200 systems. Cross-layer analysis refers to the joint examination of interactions among the application, runtime, and hardware layers to understand their combined impact on performance and energy efficiency. By combining fine-grained power measurements with controlled runtime configurations, we quantify the interactions between key optimization strategies and the underlying hardware. Our results show that leveraging high-bandwidth CPU–GPU interconnects enables asynchronous optimizer offloading, reducing energy-to-solution by up to 9-14\% while maintaining comparable time-to-solution and improving throughput by approximately 5-10\%. Activation checkpointing further reduces total energy consumption by up to 13\% with minimal performance impact, while scaling across GPUs lowers per-device power by approximately 4-5\%, reflecting improved workload distribution. These findings highlight that runtime configurations optimized for performance are not necessarily optimal for energy efficiency, and that data movement plays a dominant role in determining energy behavior at scale. Together, these insights motivate a system-aware perspective when designing and deploying energy-efficient multimodal training workflows on modern architectures.

The remainder of this paper is organized as follows. Section II presents the system and runtime background, Section III describes the measurement methodology, Section IV presents the experimental results, Section V discusses key insights and practical guidelines, and Section VI concludes the paper.

\subsection{Contributions}
The contributions of this work are summarized as follows:

\begin{itemize}[leftmargin=4mm]
    \item We present a cross-layer characterization of energy and performance trade-offs in multimodal training on NVIDIA GH200 superchips, capturing interactions among hardware, runtime strategies, and workloads, with an emphasis on heterogeneous memory hierarchies and data movement.

    \item We conduct a comprehensive study of DeepSpeed framework optimizations, including optimizer offloading, activation checkpointing, and sequence parallelism, and evaluate their impact on power, energy, time-to-solution, and computational throughput across multiple model scales and system configurations.

    \item We show that asynchronous execution improves overlap between computation and data movement, leading to better energy efficiency, more stable power behavior, and improved runtime.

    \item We demonstrate that memory-centric optimizations, such as activation checkpointing and offloading, reshape execution by trading computation for reduced memory pressure and data movement, with varying benefits across model scales.

    \item We show that data movement plays a dominant role in energy behavior at scale, and that configurations optimized for performance are not necessarily optimal for energy efficiency, highlighting the need for energy-efficient communication-aware optimization.

    \item We provide practical guidelines for energy-efficient multimodal training on modern heterogeneous systems, emphasizing system-aware strategies that jointly consider computation, communication, and memory usage.
\end{itemize}

\section{System and Runtime Overview}
We present the background on the GH200 architecture,  multimodal training, and DeepSpeed optimization strategies to establish the system and runtime context for our study. 

\subsection{NVIDIA GH200 Architecture}
\label{sec:gh200}
This section provides an overview of the NVIDIA Grace Hopper Superchip (GH200) on the JUPITER supercomputer, the target platform for this study. GH200 integrates a Grace CPU and Hopper GPU via NVLink-C2C, providing up to 450 GB/s per direction (900 GB/s total). This architecture enables cache-coherent communication and a unified virtual address space. Both processors can access system memory transparently, creating a shared memory pool while retaining distinct high-performance local memory. The Grace CPU uses high-capacity LPDDR5 memory with up to 500 GB/s bandwidth, while the Hopper GPU uses HBM3 memory with up to approximately 4 TB/s bandwidth. The memory request path depends on both the memory location and the accessing processor, so that memory accesses may traverse multiple interconnects, leading to latency and bandwidth variability. Prior studies show that careful data placement is critical, as data movement across heterogeneous memory tiers can significantly impact performance and efficiency, particularly during intensive offloading or cross-device communication workloads~\cite{elster2022nvidia,luo2024benchmarking,schieffer2024harnessing}.

\begin{figure}[htbp]
  \centering
  \includegraphics[width=0.8\linewidth]{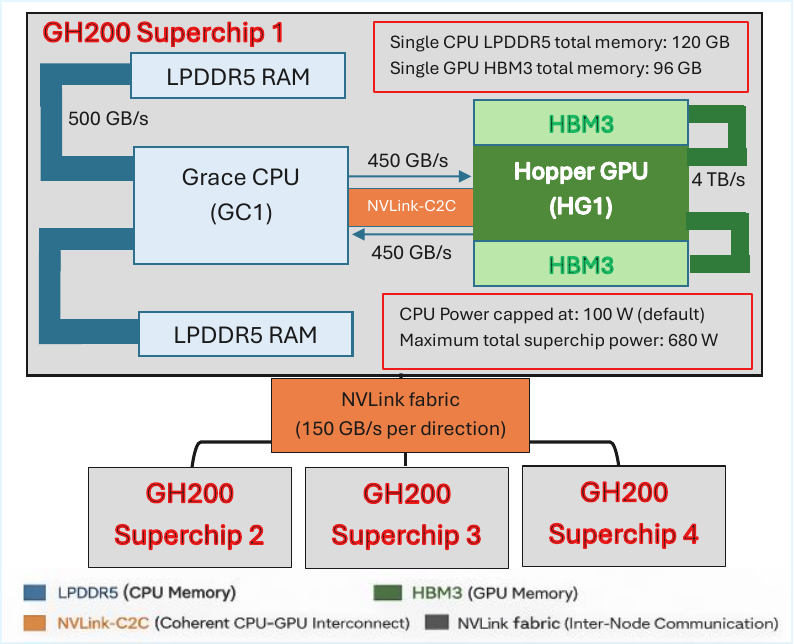}
\caption{A Cross-layer diagram shows a JUPITER compute node based on the GH200 architecture, highlighting a GH200 superchip within a quad-GPU node. It details the Grace CPU, Hopper GPU, memory hierarchy, and NVLink-C2C interconnect.}
  \label{fig:0}
\end{figure}

The power consumption of a single GH200 Superchip is determined by its integrated subsystems, including the Hopper GPU, Grace CPU, HBM and LPDDR memory, NVLink-C2C interconnect, and unified memory hardware. Although the GH200 is designed for a combined power envelope of approximately 900 to 1000 W, on JUPITER, it is limited to about 680 W to enable the superchip to operate more energy-efficiently. The actual power distribution between the CPU and GPU varies with workload and system-level power management policies. By default, the CPU is prioritized. Therefore, to avoid a power shortage for the GPU, the maximum CPU power consumption can be capped. For JUPITER, the default CPU power capping is 100 W. Each JUPITER compute node contains four GH200 superchips, creating a tightly coupled multi-accelerator system. This configuration adds energy components beyond a single superchip, including intra-node GPU-GPU communication over NVLink, CPU interconnects, network interfaces, and platform-level overheads such as power delivery, cooling, and system management. Figure~\ref{fig:0} illustrates a single JUPITER node, highlighting the detailed architecture of one GPU among the four GH200 superchips.

\subsection{Multimodal Large Language Models}
\label{sec:lmm}

Large language models (LLMs) have emerged as a powerful foundation for natural language understanding, demonstrating capability in capturing complex linguistic patterns across diverse tasks~\cite{touvron2023llamaopenefficientfoundation, openai2024gpt4technicalreport}. A common approach to extending LLMs to the multimodal domain is to integrate a vision encoder, typically a Vision Transformer (ViT)~\cite{dosovitskiy2021an}, paired with a lightweight projector network that maps visual features into the LLM's token embedding space~\cite{chen2024minigptv, li2025llavaonevision}. During training, video frames are independently encoded by the ViT, projected into visual tokens, and concatenated with textual tokens before being processed by the language model backbone~\cite{zhang2025llavavideo}.

This architecture introduces a heterogeneous computational structure within a single training step. The vision encoder operates on high-dimensional spatial features with dense local attention, while the language model processes a mixed sequence of visual and textual tokens with causal attention. To handle the positional structure of video inputs, the standard 1D Rotary Position Embedding (RoPE)~\cite{Kitaev2020Reformer:} used in most LLMs is extended to 3D RoPE, encoding spatial height, width, and temporal dimensions jointly across the frame sequence~\cite{wang2024qwen2vlenhancingvisionlanguagemodels}. This allows the model to reason about spatiotemporal relationships across frames without modifying the core attention mechanism.

The resulting token sequences are substantially longer than those encountered in text-only training. A single video sample may contribute thousands of visual tokens, far outnumbering the textual supervision tokens in a typical instruction-following training pair. This token imbalance increases activation memory and communication volume, placing significant pressure on the distributed training infrastructure. To manage the growing key-value cache for long visual sequences, modern MLLMs adopt Grouped Query Attention (GQA)~\cite{ainslie2023gqa}, which reduces the number of key and value heads relative to the number of query heads, lowering the KV cache memory footprint without substantially affecting model quality. Attention computation across both visual and language components uses FlashAttention2~\cite{dao2024flashattention}, which reduces HBM pressure and increases arithmetic intensity on tensor cores.

In this work, we study this general LMM training scenario using a Qwen2.5-based language backbone~\cite{qwen2025qwen25technicalreport} with GQA, a SigLIP vision encoder~\cite{tschannen2025siglip}, and an MLP projector with 2$\times$2 spatial pooling that reduces the token count to 196 per frame, extended with 3D RoPE~\cite{wang2024qwen2vlenhancingvisionlanguagemodels,huang2026revisiting} for video inputs. Rather than committing to a single model design, we treat this as a representative instance of the general architecture described above, and our analysis targets the overall system-level behavior, power, energy, and throughput that arise from the interaction between this heterogeneous workload and the distributed training techniques.


\subsection{DeepSpeed Runtime Features}
\label{sec:deepspeed}
DeepSpeed is a systems optimization framework that enables efficient training of large deep learning models by improving scalability, memory efficiency, and throughput through a powerful suite of complementary optimization techniques such as ZeRO, offloading, activation checkpointing, and sequence parallelism~\cite{rasley2020deepspeed}.

Zero Redundancy Optimizer (ZeRO) eliminates memory redundancy in data-parallel training by partitioning model states across the data-parallel processes rather than replicating them. ZeRO comprises three stages, which additively partition model states: ZeRO-1 partitions the optimizer; ZeRO-2 partitions the gradients; and ZeRO-3 partitions the parameters. ZeRO reduces the memory footprint of model states per GPU, enabling the training of larger models~\cite{rajbhandari2021zero}.

Offloading extends the memory capacity for model states beyond HBM limitations by leveraging heterogeneous memories such as DRAM and NVMe. Standard methods such as ZeRO-Offload~\cite{ren2021zero} and ZeRO-Infinity~\cite{rajbhandari2021zero} move optimizer states, gradients, and parameters to CPU memory but stall the GPU while the CPU completes the optimizer step. Super-Offload~\cite{lian2026superoffload} removes this bottleneck by overlapping GPU computation with the CPU-side optimizer step through asynchronous data movement over the NVLink-C2C interconnect. Designed for unified memory architectures such as the GH200, Super-Offload exploits the cache-coherent CPU-GPU link to pipeline transfers alongside forward-pass execution. Throughout this work, we refer to standard offloading as \textit{synchronous offloading} and Super-Offload as \textit{asynchronous offloading}.

Activation checkpointing reduces memory usage by storing a subset of intermediate activations during the forward pass and recomputing them during backpropagation. This approach trades increased computation for lower memory consumption and is widely used to train deeper models within fixed memory budgets~\cite{zhu2024zerof,chen2016training,rajbhandari2020zero,narayanan2021efficient}. While checkpointing reduces peak memory usage, it alters execution timelines and increases computational intensity, thereby affecting performance and energy consumption. The stored activations can be further offloaded to CPU memory, either synchronously via blocking copies or asynchronously using CUDA streams for non-blocking transfers. We refer to these as synchronous and asynchronous activation checkpointing, respectively.

For long-context and multimodal workloads, DeepSpeed offers sequence parallelism via systems such as DeepSpeed-Ulysses~\cite{jacobs2023deepspeed,yao2024ulysses}. This approach partitions input sequences across devices and uses collective communication, e.g., all-to-all exchanges, to compute attention in parallel across sequence shards. Sequence parallelism improves scalability for long sequences, maintains communication efficiency, and enables training with longer context lengths and higher throughput.

Together, ZeRO partitioning, offloading, activation checkpointing, and sequence parallelism reshape the distribution of computation, memory access, and communication during training. By dynamically moving model states and intermediate data among HBM, LPDDR, and interconnect pathways, DeepSpeed supports unprecedented model and sequence-length scales and fundamentally alters memory traffic patterns. As a result, its interaction with heterogeneous hardware architectures is critical for both performance and energy efficiency.


\section{Measurement Methodology}

This section outlines the methodology for evaluating performance and energy in multimodal training across runtime configurations. It combines fine-grained power measurements with metrics to assess throughput, time, energy, and energy efficiency. The subsections cover system setup, workloads, instrumentation, and metrics.

\subsection{Experimental Platform}

All experiments were conducted on the JUPITER supercomputer (Booster module) at the J\"ulich Supercomputing Centre (JSC), Germany, ranked fourth globally on the November 2025 TOP500 list~\cite{top500}, whose compute nodes are based on the GH200 architecture described in Section~\ref{sec:gh200}. Each node contains four GH200 superchips and is connected via the NVIDIA Quantum-2 InfiniBand, which supports high-bandwidth and low-latency communication for distributed workloads. The Booster module of JUPITER has $6{,}000$ compute nodes in $125$ racks and with around $24{,}000$ GH200 superchips.

The experimental framework is built on DeepSpeed~\cite{rasley2020deepspeed}, using the runtime optimizations described in Section~\ref{sec:deepspeed}. For asynchronous CPU offloading over the GH200 NVLink-C2C interconnect, we use SuperOffload~\cite{lian2026superoffload}. The model architecture follows the LMM configuration described in Section~\ref{sec:lmm}, evaluated at three parameter scales: 7B, 32B, and 72B. We use a subset from LLaVA-Video-178K~\cite{lin2023videollava} for our experiments.

Our experiments cover single- and multi-node setups. The single-node configuration isolates intra-node behavior (CPU–GPU interaction, memory, and kernel-level energy), while the multi-node setup evaluates scalability, communication overhead, and total energy. For each case, we maximize local batch size within HBM limits and use activation checkpointing for long contexts. Experiments use the minimum number of nodes required to support the model, with the local batch size maximized to fully utilize available HBM at each configuration. The number of training samples in each group is kept constant to ensure equivalent total FLOPs across groups.

We measure power and energy for each training run using a dual-channel monitoring approach. Board-level power is sampled from the Linux \texttt{hwmon} subsystem via the \texttt{power1\_average} sysfs interface at 100 ms intervals, collecting data from all sixteen per-socket sensors on the four-socket GH200 node. These sensors include Module Power, Grace CPU Power, CPU Core Power, and SysIO (I/O and memory) Power, with one instance per socket.

Module Power, measured at the voltage regulator input, captures the total consumption of all on-package components, including Grace CPU cores, Hopper GPU die, HBM3, LPDDR5X, and NVLink-C2C overhead. This serves as our primary node-level energy metric. GPU die power is also sampled using NVML at the same interval, providing a per-chip perspective that complements board-level measurements without double-counting. Moreover, all power streams are synchronized using Unix timestamps and correlated with training phase markers (such as forward pass, visual encoding, LLM execution, and backward pass) injected into the training loop.

\subsection{Workloads and Configurations}

We organize experiments into multiple configurations to systematically study the effects of runtime strategies and system scaling. For optimizer offloading, configurations A1 to A12 are defined for each model scale: A1, A5, and A9 are baseline runs without offloading; A2, A6, and A10 use scaled GPU configurations without offloading; A3, A7, and A11 use synchronous CPU offloading; and A4, A8, and A12 use asynchronous CPU offloading. For activation offloading, configurations A1 to A9 are used: A1, A4, and A7 represent no-offloading; A2, A5, and A8 use synchronous offloading; and A3, A6, and A9 use asynchronous offloading. Sequence parallelism is evaluated at three levels (1, 2, 4) under fixed- and scaled-GPU configurations to examine the trade-off between parallel efficiency and communication overhead. All configurations use fixed batch sizes, sequence lengths, and data preprocessing pipelines to ensure fair comparisons.

These controlled variations enable an analysis of how runtime strategies and parallelism interact with system architecture to affect performance, time, and energy consumption.

\subsection{Metrics}
We evaluate runtime strategies using metrics that capture time, energy, energy efficiency, and computational throughput.

{\bf Training Time: }
We measure total wall-clock training time, capturing end-to-end performance including computation, communication, and data movement overheads.

{\bf Throughput: }
Throughput (TFLOP/s) is computed as the total FP operations (ViT and LLM) divided by the execution time, thereby measuring hardware utilization.

{\bf Average Power: }
Average power consumption (in Watts) is computed as the mean power usage over the full execution: $P_{\text{avg}} = \frac{1}{T} \int_{0}^{T} P(t)\, dt,$ where $P(t)$ is the instantaneous power and $T$ is the total runtime. We report power at both the component level (CPU, GPU) and system level (superchip).

{\bf Energy: }
Energy (in kiloJoules) represents the total energy consumed during the execution period (one or more training steps) and is defined as: $E = \int_{0}^{T} P(t)\, dt. $


{\bf Energy Efficiency: }
Energy efficiency is defined as the ratio of computation to energy consumed: $\text{Efficiency} = \frac{\text{FLOPs}}{E} $, showing how effectively energy is converted into computation.

\section{Experiments}
This section presents a cross-layer analysis of power, energy, and performance for multimodal training on GH200 systems using DeepSpeed. We evaluate key runtime strategies, including CPU offloading and activation checkpointing, and compare no-offloading, synchronous, and asynchronous execution across model scales (7B, 32B, 72B) in single-node and distributed settings. The objective is to quantify the impact of data movement and runtime design on efficiency and to identify configurations that balance energy and performance.

\subsection{Optimizer Offloading}

We evaluate the impact of DeepSpeed optimizer CPU offloading on power distribution and energy efficiency in multimodal training. In video-based multimodal models, the CPU primarily handles data ingestion tasks, including video frame loading and decoding, while the GPU performs the majority of the computation. Enabling optimizer offloading moves optimizer states from GPU HBM to CPU memory, thereby alleviating GPU memory pressure and enabling larger model configurations or larger batch sizes. From a cross-layer perspective, this redistribution of data and computation alters memory access patterns, increases CPU–GPU data movement over the interconnect, and reshapes device utilization. To capture these effects, we analyze the behavior of the multimodal model across three scales (7B, 32B, and 72B) under optimizer offloading, focusing on trade-offs among power, data movement, energy, and performance.

For the 7B model, experiments are conducted on 4 and 8 GPUs; for the 32B model on 16 and 32 GPUs; and for the 72B model on 32 and 64 GPUs. Increasing the number of GPUs enables larger effective batch sizes and improves throughput under distributed execution without offloading. For each model size, configurations A1, A5, and A9 correspond to baseline runs without offloading, using a fixed number of GPUs, whereas A2, A6, and A10 represent no-offloading runs with 2$\times$ GPU scaling to increase the workload and expose distributed execution behavior. Offloading strategies are evaluated independently at the baseline scale, where A3, A7, and A11 correspond to synchronous CPU offloading, and A4, A8, and A12 correspond to asynchronous CPU offloading, both using the same number of GPUs as the baseline configuration. All experiments are conducted with a total of 1.6k training iterations, using local batch sizes of 4, 2, and 1 for 7B, 32B, and 72B, respectively, for the baseline. For the remaining cases, we increase it by a factor of 2. This setup isolates the effects of offloading from those of system scaling, enabling a clear comparison of how runtime strategies impact power, energy, energy efficiency, time, and TFLOP/s.

\begin{figure}[htbp]
  \centering
  \begin{subfigure}[t]{0.49\linewidth}
    \centering
    \includegraphics[width=\linewidth]{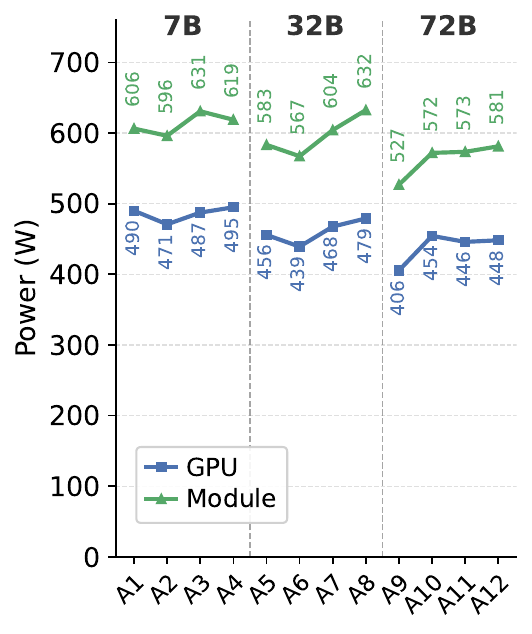}
    \caption{Average power breakdown \\(GPU and total module).}
    \label{fig:1}
  \end{subfigure}
  \hfill
  \begin{subfigure}[t]{0.49\linewidth}
    \centering
    \includegraphics[width=\linewidth]{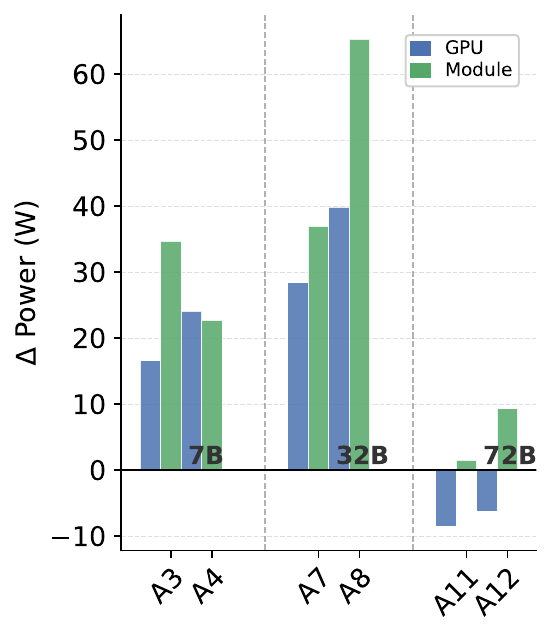}
    \caption{Power variation with offloading relative to scaled configurations.}
    \label{fig:11}
  \end{subfigure}

  \caption{Power behavior across configurations and model scales (7B, 32B, and 72B). Baseline configurations (A1/A5/A9) use no-offloading, scaled configurations (A2/A6/A10) increase GPU count, and A3-A12 apply synchronous/asynchronous CPU offloading.} 
  \label{fig:111}
\end{figure}

{\bf Power and Energy: }
SubFigure~\ref{fig:1} presents the average power across configurations, including both GPU and total module power, while SubFigure~\ref{fig:11} illustrates the power variation relative to the scaled configurations. Within each model group (7B: A1-A4, 32B: A5-A8, 72B: A9-A12), the progression from no-offloading to synchronous and asynchronous offloading highlights how runtime strategies reshape power behavior. Across all configurations, GPU power remains the dominant contributor to total module power, while CPU power ranges from 55 to 71 W. The module power corresponds to the GH200 superchip consumption measured at the board level, including the GPU, CPU, HBM, LPDDR, NVLink-C2C interconnect, and associated on-package components.
 
A comparison between the baseline configurations (A1, A5, A9) and their scaled counterparts (A2, A6, A10) shows that increasing the number of GPUs leads to a modest reduction in per-GPU power for the 7B and 32B models. In the 7B case, per-GPU power decreases from about 490 W to 471 W, indicating improved workload distribution and reduced computational pressure on individual devices. In the 32B model, the power variations are approximately 17 W. This trend becomes less pronounced for the 72B model, where per-GPU power increases by about 48 W as scaling proceeds. This suggests that, at larger model sizes, GPUs operate closer to saturation, and additional parallelism reduces per-device load. The impact of communication overheads and synchronization costs becomes more significant, partially offsetting the gains from workload distribution.

When offloading is introduced at the baseline scale, per-GPU power increases moderately, rising from 490 W to 495 W (+5 W) for the 7B model, from 456 W to 479 W (+23 W) for the 32B model, and from 406 W to 448 W (+42 W) for the 72B model (asynchronous case). This increase reflects improved GPU utilization, driven by reduced memory stalls and enhanced overlap between computation and data movement. Module power demonstrates a comparable trend, increasing during offloading (e.g., reaching up to 479 W for 32B). Asynchronous offloading consistently yields slightly higher and more stable power levels than synchronous execution, suggesting improved overlap and smoother execution dynamics. These findings show that offloading shifts the system toward higher utilization rather than reducing instantaneous power, which is a critical factor in enhancing energy efficiency. Moreover, for the 7B and 32B models, offloading increases both GPU and module power compared to the scaled configurations (A2 and A6), reflecting improved GPU utilization due to reduced memory stalls and better overlap between computation and data movement. In contrast, for the 72B model, offloading slightly reduces GPU power while increasing total module power. This indicates a shift in activity from the GPU to other components (CPU, memory, and interconnect), as GPUs are already near saturation, and data movement and synchronization overheads dominate.

\begin{figure}[htbp]
  \centering
  \begin{subfigure}[t]{0.49\linewidth}
    \centering
    \includegraphics[width=\linewidth]{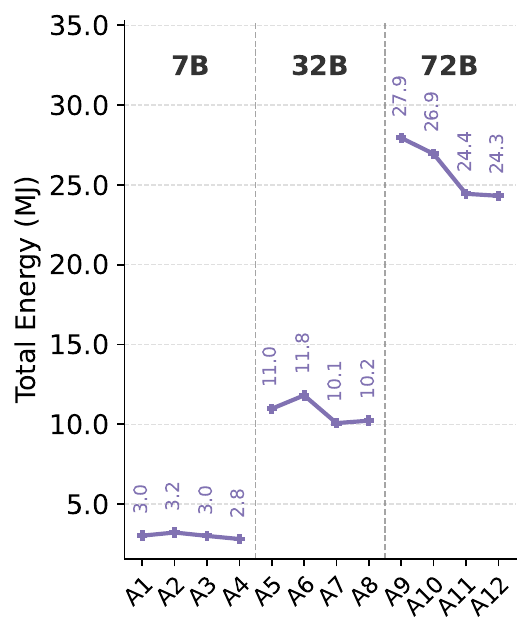}
    \caption{Total energy consumption (kJ).}
    \label{fig:2}
  \end{subfigure}
  \hfill
  \begin{subfigure}[t]{0.49\linewidth}
    \centering
    \includegraphics[width=\linewidth]{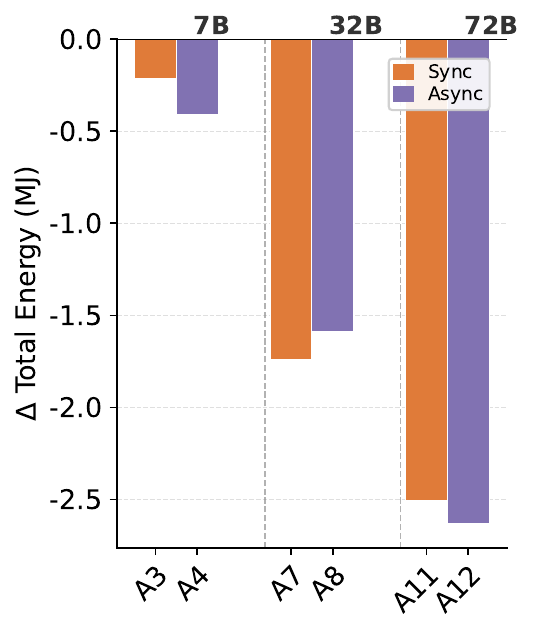}
    \caption{Energy variation with offloading relative to scaled configurations.}
    \label{fig:3}
  \end{subfigure}

  \caption{Total energy consumption and relative energy variation across configurations (A1-A12) and model scales (7B, 32B, and 72B).}
  \label{fig:23}
\end{figure}

Figure~\ref{fig:2} and Figure~\ref{fig:3} show the total energy consumption of a single training step and the relative energy change compared to the no-offloading baseline across all configurations. For each model scale, offloading reduces energy compared to the no-offloading baseline. In the 7B model, energy decreases from approximately 3020 kJ (A1) to 2810 kJ (A4). In the 32B model, the energy decreases from approximately 10972 kJ (A5) to about 10224 kJ (A8). For the 72B model, energy decreases significantly from approximately 27943 kJ (A9) to around 24314 kJ (A12), despite the larger workload. Asynchronous offloading achieves the greatest reductions, saving approximately 12.72\% for 7B, 13.45\% for 32B, and 9.76\% for 72B compared to the baseline configuration. Synchronous offloading also reduces energy consumption, though to a lesser extent, with savings of 6.61 \%, 14.71\%, and 9.29\% for 7B, 32B, and 72B, respectively. These results demonstrate that offloading effectively reduces energy consumption while maintaining workload. The small difference between synchronous and asynchronous modes is due to the high-bandwidth NVLink-C2C interconnect of GH200, which enables efficient overlap of data movement and GPU computation. Across all model sizes, scaled configurations do not achieve energy savings compared to offloading.

\begin{figure*}[t]
  \centering
  \includegraphics[width=0.75\linewidth]{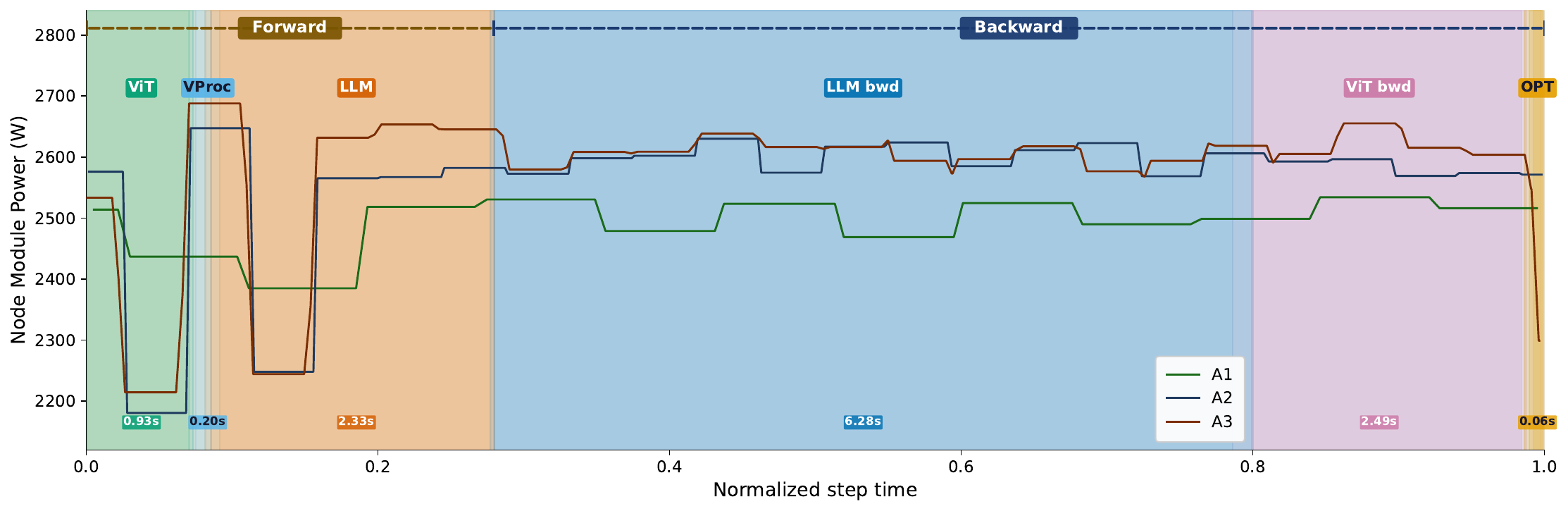}
  \caption{Node-level power profile over a normalized training step for the 7B model under baseline (A1), synchronous (A3), and asynchronous (A4) offloading configurations. The timeline separates forward/backward phases, with distinct ViT, LLM, and optimizer regions.}
  \label{fig:power_signal}
\end{figure*}

Figure~\ref{fig:power_signal} presents the node-level power profile over a complete training step, demonstrating that the baseline, synchronous, and asynchronous offloading configurations exhibit similar overall trends but differ in the consistency of power maintenance across phases. Overall, offloading configurations (A2 and A3) exhibit higher power consumption than the baseline configuration (A1), primarily due to the increased workload and larger batch sizes enabled by offloading optimizer states to CPU memory. During the forward pass, particularly in the ViT and LLM regions, the power profile exhibits noticeable fluctuations driven by workload and batch size, which are lower in A1 compared to A2 and A3. Baseline configuration (A1) exhibits more stable power behavior. Synchronous offloading exhibits slightly more pronounced power drops than the asynchronous variant. Conversely, in the backward pass, asynchronous offloading maintains a more stable, consistently higher power level, reflecting improved overlap between computation and data movement. This approach reduces idle periods and minimizes power dips compared to the synchronous configuration, which introduces brief stalls due to periodic synchronization. At the OPT stage, the power in A3 (asynchronous) drops, as data movement has already been overlapped with computation in previous stages, unlike in A1 (baseline) and A2 (synchronous). In summary, asynchronous offloading yields a smoother power profile and improves resource utilization, particularly during the backward phase, which accounts for the majority of the computation.

\begin{figure}[t]
  \centering
  \begin{subfigure}[t]{0.3\linewidth}
    \centering
    \includegraphics[width=\linewidth]{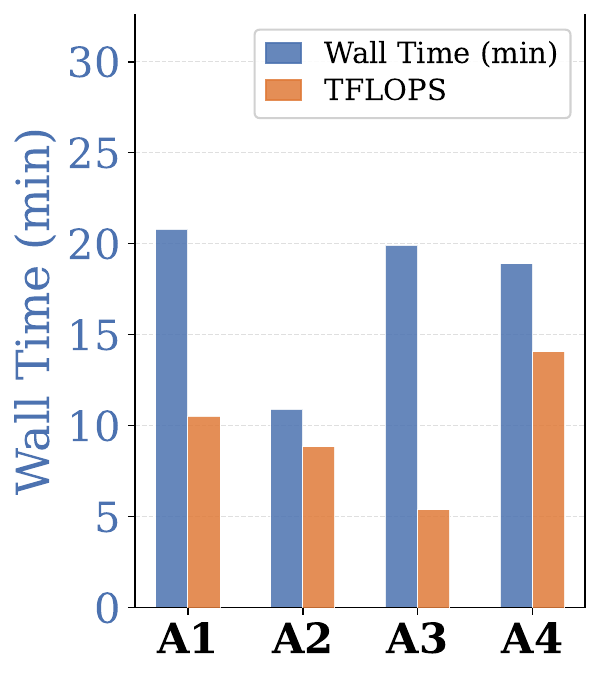}
    \caption{7B}
  \end{subfigure}
  \hfill
  \begin{subfigure}[t]{0.3\linewidth}
    \centering
    \includegraphics[width=\linewidth]{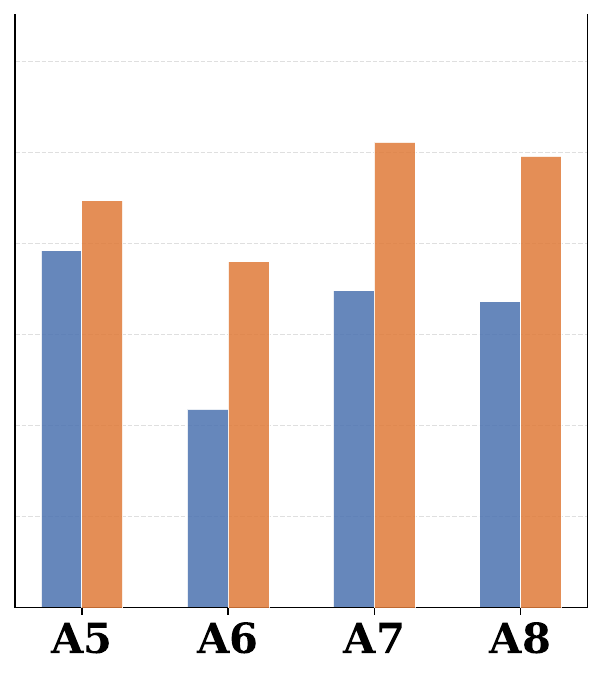}
    \caption{32B}
  \end{subfigure}
  \hfill
  \begin{subfigure}[t]{0.3\linewidth}
    \centering
    \includegraphics[width=\linewidth]{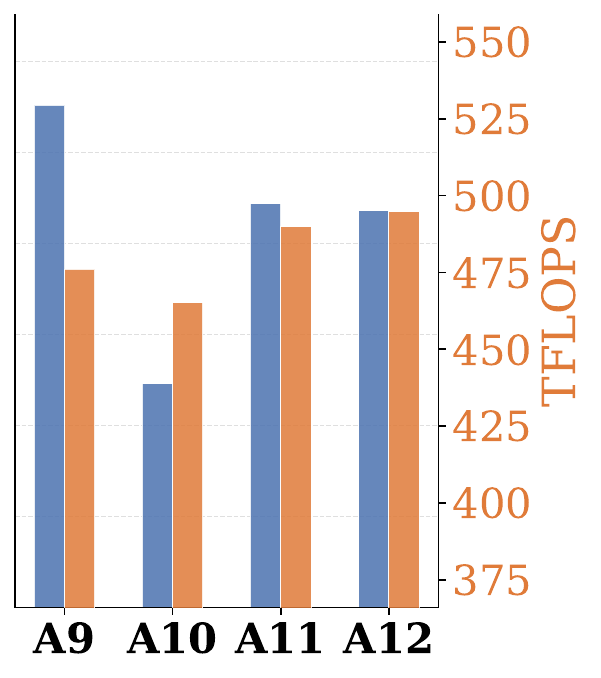}
    \caption{72B}
  \end{subfigure}

  \caption{Performance trade-off between time-to-solution (min) and throughput (TFLOP/s) across model scales.} 

  \label{fig:4}
\end{figure}

{\bf Throughput and Time-to-Solution:}
Figure~\ref{fig:4} shows that offloading consistently improves both throughput and time-to-solution (Wall Time) across model scales. For the 7B model, TFLOP/s increases from 428.21 (baseline) to 449.16 (asynchronous), while wall time decreases from 20.8 minutes to 18.9 minutes. For the 32B model, throughput increases from 478.39 TFLOP/s to over 512 TFLOP/s, and wall time decreases from 19.6 minutes to 16.8 minutes under asynchronous execution. For the 72B model, TFLOP/s increases from 465.11 to 494.72, while wall time decreases from 27.6 minutes to 21.8 minutes. As illustrated in Figure~\ref{fig:4}, asynchronous offloading provides higher throughput and lower execution time. It provides the best trade-off between performance and efficiency, enabling improved GPU utilization without increasing time.

\subsection{Activation Offloading}

\begin{figure}[htbp]
  \centering
  \includegraphics[width=0.85\linewidth]{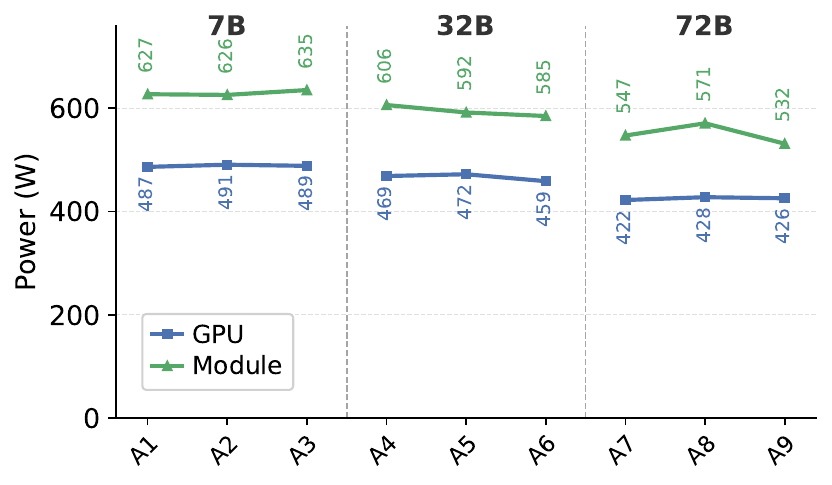}
  \caption{Average power breakdown (GPU and total module) under activation offloading across configurations (A1-A9), grouped by model scales (7B, 32B, 72B).} 
  \label{fig:5}
\end{figure}

{\bf Power and Energy Characterization:}
Figure~\ref{fig:5} presents the average GPU and module power when enabling DeepSpeed activation offloading across different model scales. GPU power remains the primary contributor in all configurations. For the 7B model, GPU power is nearly constant across configurations (486.6 W for baseline, 490.8 W for synchronous, and 488.6 W for asynchronous offloading), indicating that activation offloading does not significantly affect GPU utilization at this scale. Module power varies slightly (627.3 W to 635.4 W), reflecting modest increases in CPU and memory activity. In the 32B model, asynchronous offloading results in a slight reduction in GPU power (468.9 W to 458.9 W), indicating improved execution efficiency and reduced memory pressure. Module power also decreases from 606.4 W (baseline) to 585.1 W (async), indicating reduced overall system activity at this scale. For the 72B model, GPU power remains relatively stable (422.4 W to 427.9 W), while module power varies more substantially (547.3 W to 571 W). This variation reflects increased CPU and interconnect activity required to support activation movement, although GPU utilization remains largely unchanged. Overall, activation offloading has a limited effect on GPU power, with most variation occurring at the module level due to increased data movement and CPU involvement.

\begin{figure}[t]
  \centering
  \begin{subfigure}[t]{0.47\linewidth}
    \centering
    \includegraphics[width=\linewidth]{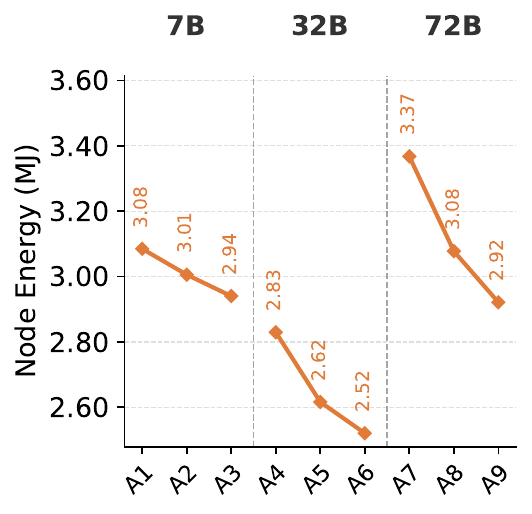}
    \caption{Total node-level energy (MJ).}
    \label{fig:7a}
  \end{subfigure}
  \hfill
  \begin{subfigure}[t]{0.47\linewidth}
    \centering
    \includegraphics[width=\linewidth]{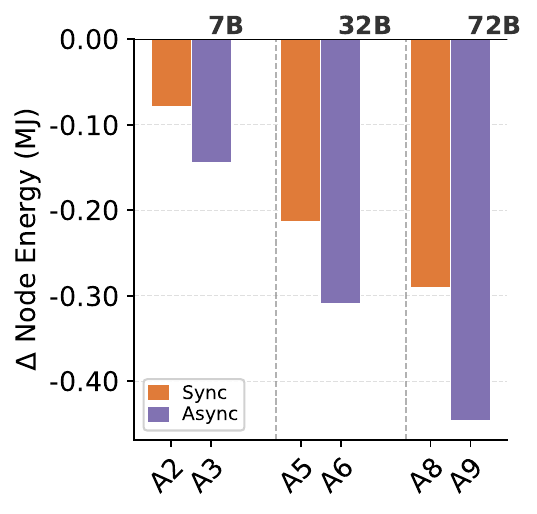}
    \caption{Energy reduction relative to baseline.}
    \label{fig:7b}
  \end{subfigure}

  \caption{Node-level energy behavior under activation checkpointing across configurations (A1-A9) and model scales (7B, 32B, and 72B).}
  \label{fig:56}
\end{figure}

SubFigure~\ref{fig:7a} reports total node-level energy consumption across configurations, and SubFigure~\ref{fig:7b} highlights the relative energy savings compared to the no-offloading baseline. As shown, activation offloading consistently reduces energy per training step across all model sizes. For the 7B model, energy consumption decreases from 3084.8 kJ (baseline) to 3005.7 kJ (sync) and 2940.2 kJ (async), corresponding to savings of approximately 2.6\% and 4.7\%, respectively. For the 32B model, energy consumption is reduced from 2829.7 kJ to 2616.2 kJ (sync) and 2520.7 kJ (async), yielding savings of approximately 7.5\% and 10.9\%. For the 72B model, activation offloading yields the largest gains, reducing energy consumption from 3367.6 kJ to 3077.6 kJ (sync) and 2921.3 kJ (async), corresponding to savings of approximately 8.6\% and 13.2\%, respectively. These results show that, unlike power, energy savings grow with model size. Asynchronous activation offloading yields the largest gains by overlapping data movement with computation, thereby reducing idle time.

{\bf Throughput and Time-to-Solution:}
Activation offloading improves both throughput and time-to-solution across all model scales, with the most consistent gains observed under asynchronous execution. For the 7B model, TFLOP/s increases from 434.15 to 446.35 (+2.8\%), while wall time decreases from 20.5 to 19.3 minutes (-5.9\%). For the 32B model, asynchronous offloading maintains comparable throughput (500.5 vs.\ 502.85 TFLOP/s, -0.5\%) while reducing wall time from 19.4 to 18.0 minutes (-7.2\%). 
For the 72B model, the improvements are more pronounced, with TFLOP/s increasing from 488.91 to 523.07 (+7.0\%) and wall time decreasing from 25.6 to 22.9 minutes (-10.5\%). Overall, asynchronous activation offloading provides the best balance between performance and execution time, enabling higher effective throughput while reducing time-to-solution, particularly for larger models.

\subsection{Sequence Parallelism Degree}

We assess sequence parallelism (SP=1, 2, 4) under fixed and scaled GPU configurations. 

\begin{figure}[htbp]
  \centering
  \begin{subfigure}[t]{0.32\linewidth}
    \centering
    \includegraphics[width=\linewidth]{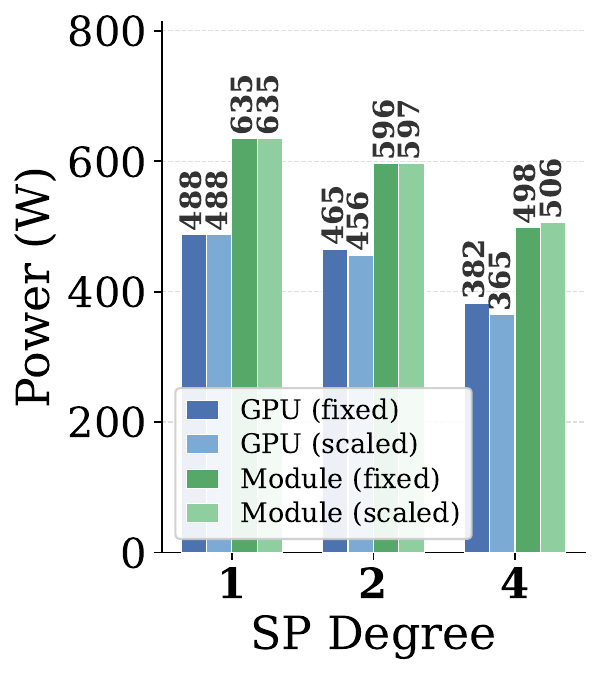}
    \caption{7B}
  \end{subfigure}
  \hfill
  \begin{subfigure}[t]{0.32\linewidth}
    \centering
    \includegraphics[width=\linewidth]{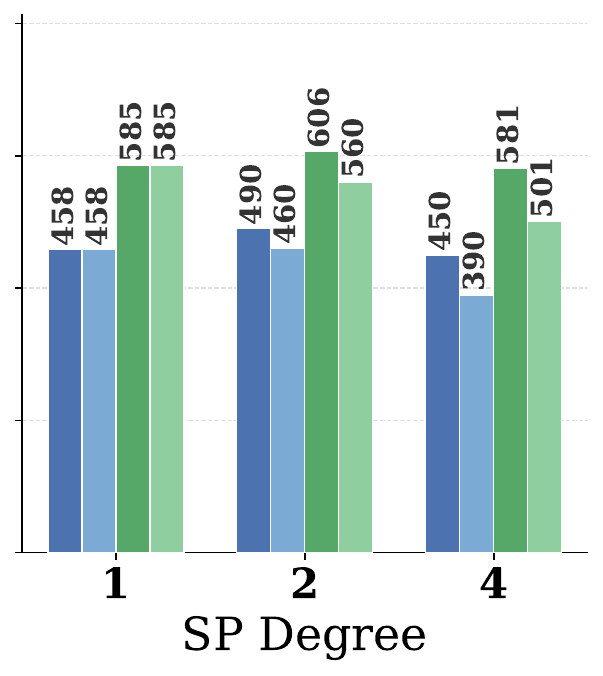}
    \caption{32B}
  \end{subfigure}
  \hfill
  \begin{subfigure}[t]{0.32\linewidth}
    \centering
    \includegraphics[width=\linewidth]{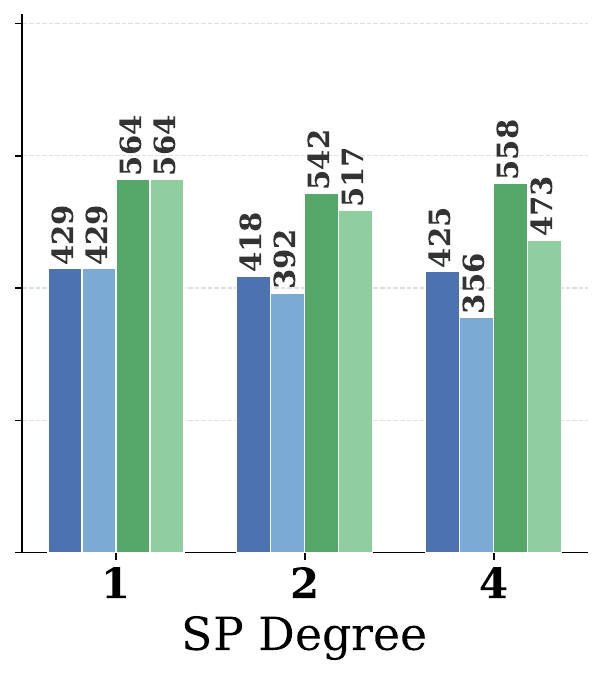}
    \caption{72B}
  \end{subfigure}

  \caption{Average GPU and module power across SP degrees for different model scales under fixed and scaled GPU configurations.}

  \label{fig:7}
\end{figure}

\begin{figure}[!htbp]
  \centering
  \begin{subfigure}[t]{0.31\linewidth}
    \centering
    \includegraphics[width=\linewidth]{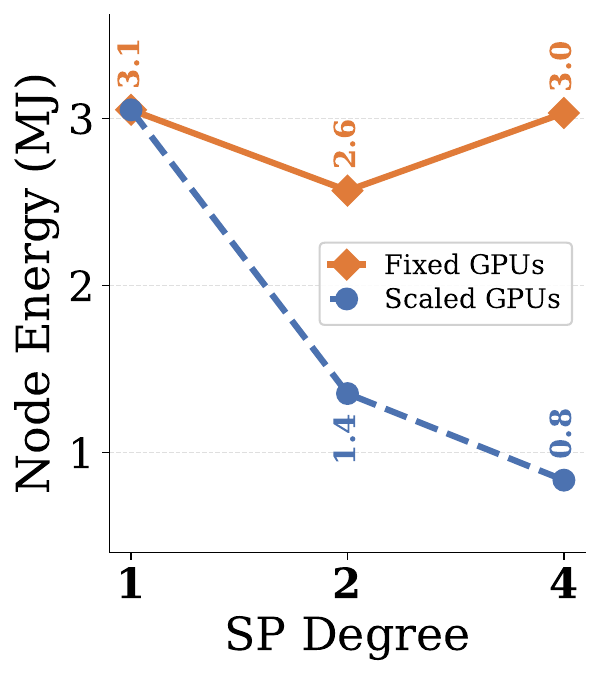}
    \caption{7B}
  \end{subfigure}
  \hfill
  \begin{subfigure}[t]{0.31\linewidth}
    \centering
    \includegraphics[width=\linewidth]{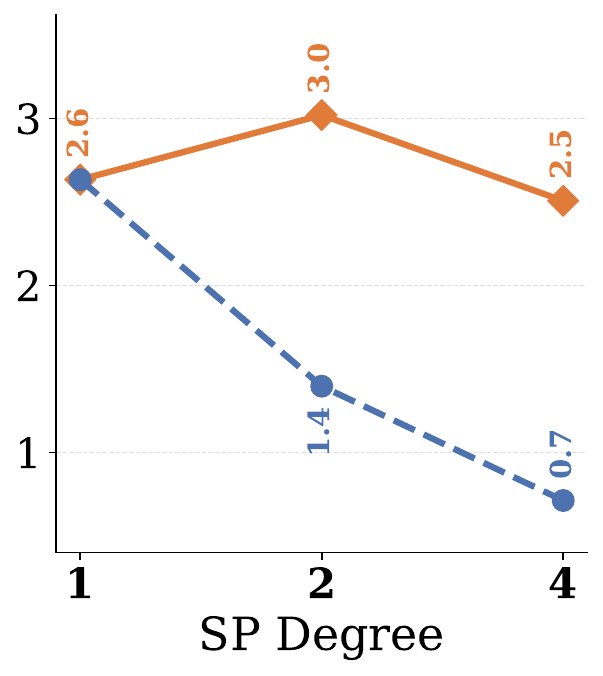}
    \caption{32B}
  \end{subfigure}
  \hfill
  \begin{subfigure}[t]{0.31\linewidth}
    \centering
    \includegraphics[width=\linewidth]{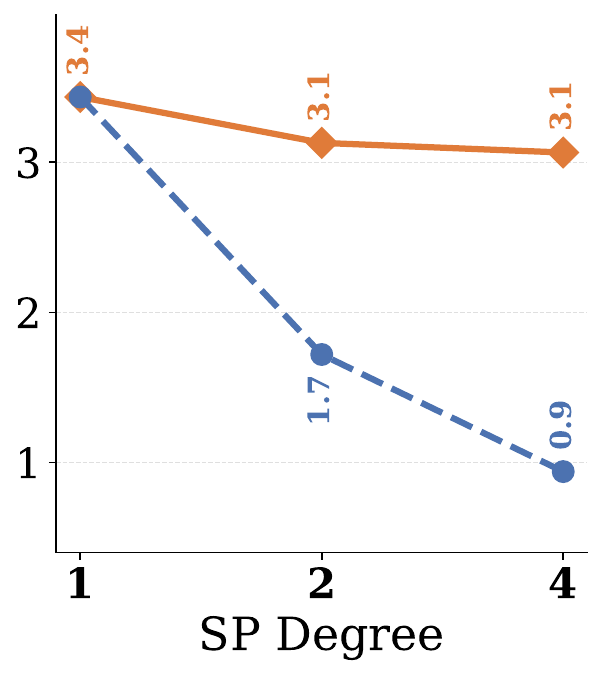}
    \caption{72B}
  \end{subfigure}

  \caption{Node-level energy consumption (MJ) across sequence parallelism levels (SP=1, 2, 4) for model sizes (7B, 32B, 72B).}

  \label{fig:8}
\end{figure}

\begin{figure}[!htbp]
  \centering
  \begin{subfigure}[t]{0.31\linewidth}
    \centering
    \includegraphics[width=\linewidth]{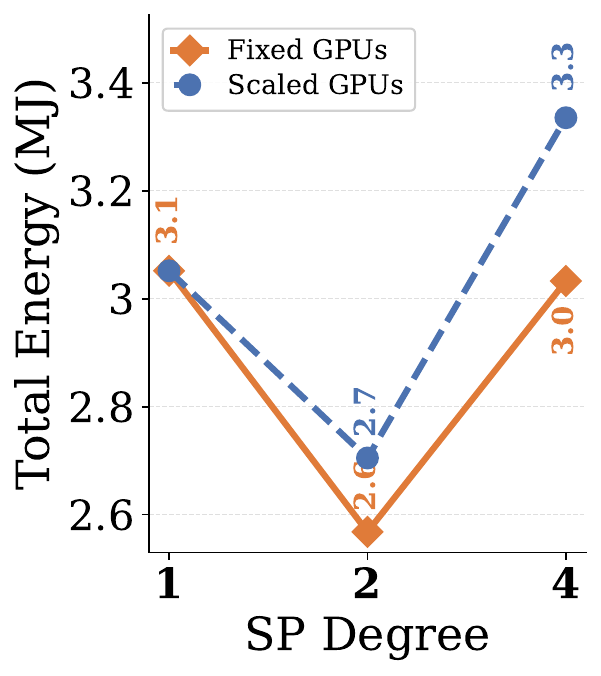}
    \caption{7B}
  \end{subfigure}
  \hfill
  \begin{subfigure}[t]{0.31\linewidth}
    \centering
    \includegraphics[width=\linewidth]{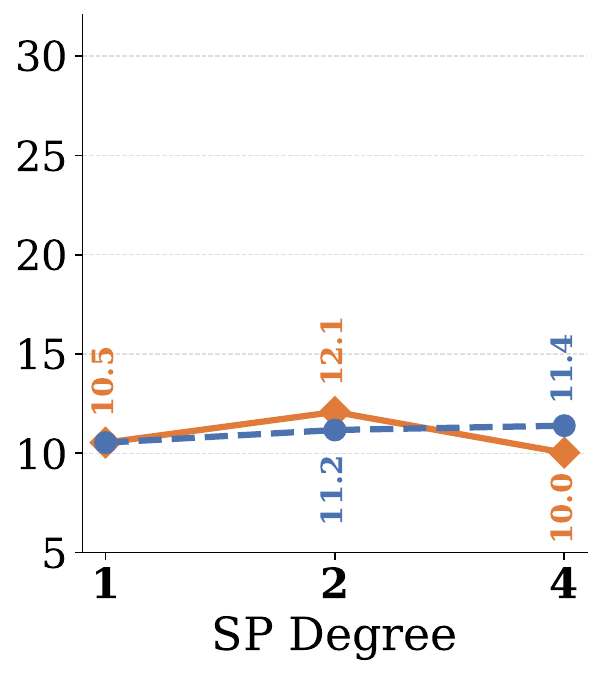}
    \caption{32B}
  \end{subfigure}
  \hfill
  \begin{subfigure}[t]{0.31\linewidth}
    \centering
    \includegraphics[width=\linewidth]{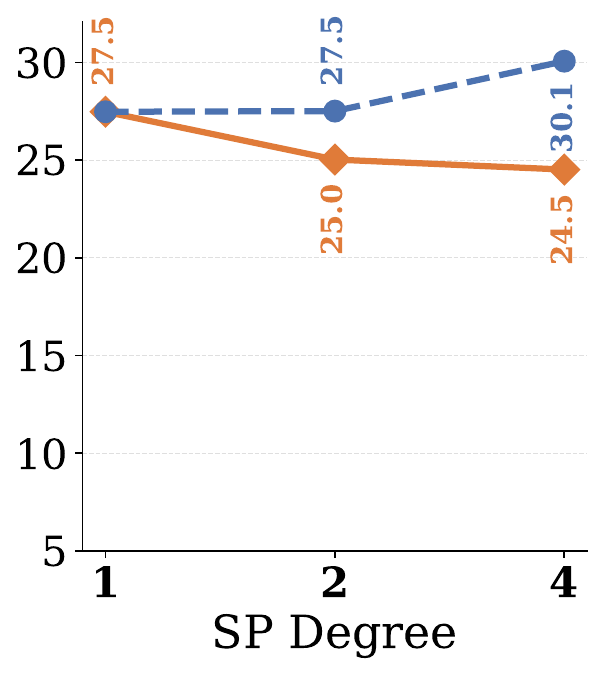}
    \caption{72B}
  \end{subfigure}

  \caption{Total energy consumption (kJ) across sequence parallelism degrees (SP=1, SP=2, SP=4) for model scales (7B, 32B, 72B).}

  \label{fig:9}
\end{figure}

\textbf{Power and Energy:} 
Figure~\ref{fig:7} shows the GPU and module power across SP degrees, while Figures~\ref{fig:8} and~\ref{fig:9} report node-level and total energy consumption, respectively. As observed, increasing the SP degree generally reduces GPU power for smaller models due to improved workload partitioning. For the 7B model, GPU power drops from 488.0 W (SP=1) to 382.3 W (SP=4), and module power decreases from 635.2 W to 498.4 W. Node energy falls to 2568.1 kJ at SP=2 from 3051.8 kJ at SP=1, though excessive partitioning at SP=4 slightly increases total energy due to the all-to-all communication overhead. For the 32B model, power trends are less consistent. GPU power peaks at SP=2 (490.0 W) and drops at SP=4 (449.7 W), while module power ranges from 585.4 W to 606.3 W. SP=4 yields the lowest node energy (2507.1 kJ), reflecting improved efficiency despite moderate power variation. For the 72B model, GPU power remains relatively stable across SP degrees (429.0 W at SP=1, 417.9 W at SP=2, 425.2 W at SP=4), and module power follows a similar pattern (564.2 W, 542.5 W, 558.0 W). Under the fixed-GPU configuration, node energy decreases steadily from 3434.4 kJ at SP=1 to 3128.8 kJ at SP=2 and 3064.5 kJ at SP=4, showing consistent but moderate gains from sequence partitioning. Under the scaled-GPU configuration, the reduction is far more pronounced, with node energy dropping from 3434.4 kJ to 1719.4 kJ (64 GPUs, SP=2) and 939.5 kJ (128 GPUs, SP=4), demonstrating that combining higher SP degrees with additional GPUs is especially effective at reducing energy-to-solution for large models. Overall, increasing SP reduces energy across all models, with the largest gains for 72B under scaled setups, though diminishing returns and communication overheads become more pronounced for smaller models.


\textbf{Throughput and Time-to-Solution:}
Increasing the SP degree improves time-to-solution under scaled configurations, but introduces trade-offs in throughput. For the 7B model, wall time decreases from 20.0 minutes (SP=1) to 6.9 minutes (SP=4 with scaling), while TFLOP/s slightly increases from 429.01 to 449.72. For the 32B model, wall time improves significantly from 18.8 minutes to 5.9 minutes under scaling, although TFLOP/s decreases from 495.91 to 460.24 due to increased communication overhead. For the 72B model, the largest gains occur, with wall time dropping from 25.4 minutes (SP=1) to 8.3 minutes (SP=4 with scaling), while TFLOP/s decreases from 476.03 to 419.37. Higher SP reduces runtime, especially with GPU scaling, but lowers per-device efficiency. Overall, it improves scalability and reduces time/energy with moderate communication overhead.

\subsection{ZeRO Partitioning}
We evaluate ZeRO stages 1, 2, and 3 on a 7B model across 4, 16, 32, and 64 GPUs to examine the trade-off between communication overhead, memory savings, and energy efficiency. ZeRO-3's deeper parameter partitioning frees sufficient GPU memory to double the sequence length (4x12k) compared to ZeRO-1 and ZeRO-2 (2x12k), allowing us to assess whether the additional communication cost is offset by improved utilization.
At small GPU counts, shallower partitioning holds a slight energy advantage. With 4 GPUs, all stages yield similar GPU power (483–492 W) and wall time (79–82 minutes), but ZeRO-1 achieves the lowest node energy (11,569 kJ) due to minimal communication overhead. At 16 GPUs, this gap persists: ZeRO-1 and ZeRO-2 consume approximately 3,230–3,240 kJ, while ZeRO-3 remains comparable at 3,163 kJ despite processing the larger batch.
As the GPU count increases, the efficiency gap between stages narrows. At 32 GPUs, wall times converge to 11.6–12.4 minutes and node energy ranges from 1,643 kJ (ZeRO-1) to 1,740 kJ (ZeRO-2), with ZeRO-3 at 1,661 kJ. At 64 GPUs, the differences become negligible: wall times range from 6.4 to 6.8 minutes across all stages, and node energy varies by less than 6\% (825–871 kJ). This convergence indicates that the communication overhead of deeper partitioning is increasingly amortized as devices scale.
Overall, the energy cost of deeper ZeRO partitioning diminishes with GPU scaling, such that all three stages achieve comparable power and energy efficiency at 64 GPUs. Deeper partitioning frees memory for larger batches and longer sequences, improving utilization and throughput without a proportional increase in energy cost. Thus, ZeRO stage choice is driven by memory and workload, not energy.


\subsection{Energy Efficiency}
Energy efficiency reflects how effectively energy is converted into useful computation, but it can degrade significantly when excessive data movement is involved. Table~\ref{tab:energy_efficiency} presents overall run efficiency (TFLOP/kJ) for each technique and model scale, with gains shown relative to the baseline. 

Optimizer offloading provides limited and inconsistent benefits. Synchronous offloading reduces efficiency for the 7B model ($-6.4\%$) but yields small gains for the 32B ($+0.9\%$) and 72B ($+1.1\%$) models. Asynchronous offloading improves efficiency for the 7B model ($+7.5\%$) but decreases it for 32B ($-4.8\%$) while remaining near-neutral for 72B ($+0.8\%$). These results suggest that the data movement overhead introduced by optimizer offloading can outweigh its benefits, with the impact depending on model scale and the degree of compute-communication overlap.

Activation checkpointing offers consistent improvements, particularly in asynchronous mode. Synchronous checkpointing provides modest gains for the 7B model ($+1.9\%$) and strong improvements for 72B ($+8.0\%$), though it slightly reduces efficiency for 32B ($-1.2\%$). Asynchronous checkpointing improves efficiency across all model scales, with gains of $+4.2\%$ for 7B, $+1.0\%$ for 32B, and $+16.5\%$ for 72B. The increasing benefit at larger scales indicates that recomputation becomes more energy-efficient as memory pressure grows, reducing costly memory transfers. Compared to optimizer offloading, activation checkpointing requires less data movement and thus provides consistent energy gains.

Sequence parallelism yields the largest gains for the 7B model, with up to a $+39.5\%$ improvement at SP=4 under fixed GPU (F) configuration and a $+32.4\%$ improvement under scaled (S) configuration. For 32B, gains are modest under fixed GPUs ($+0.2\%$ at SP=4) but improve with GPU scaling ($+7.7\%$ at SP=4). For 72B, fixed configurations show mixed results ($+2.7\%$ at SP=2, $-0.8\%$ at SP=4), while scaled configurations yield consistent improvements up to $+5.3\%$. These trends show that for larger models, the communication overhead of sequence partitioning can offset its benefits unless additional GPUs are provisioned to maintain a sufficient computation-to-communication ratio. Techniques that reduce memory pressure with minimal data movement, such as asynchronous checkpointing, provide the most consistent efficiency benefits across scales. In contrast, communication-heavy approaches may reduce efficiency despite improving runtime, underscoring the need for system-aware optimizations that jointly consider performance, communication, and energy.


\begin{table}[htbp]
\centering
\caption{Energy efficiency (TFLOP/kJ) across model scales.}
\label{tab:energy_efficiency}
\setlength{\tabcolsep}{2pt}
\renewcommand{\arraystretch}{1.1} 
\small
\begin{tabular}{llccc}
\hline
Method & Config. & 7B & 32B & 72B \\
\hline
Baseline & Base & 168.7 & 212.4 & 211.3 \\
\hline
\multirow{2}{*}{Opt. Offload}
 & Sync  & 157.9 (-6.4) & 214.4 (+0.9) & 213.7 (+1.1) \\
 & Async & 181.3 (+7.5) & 202.2 (-4.8) & 212.9 (+0.8) \\
\hline
\multirow{2}{*}{Act. Ckpt.}
 & Sync  & 171.9 (+1.9) & 209.9 (-1.2) & 228.2 (+8.0) \\
 & Async & 175.8 (+4.2) & 214.5 (+1.0) & 246.1 (+16.5) \\
\hline
\multirow{4}{*}{Seq. Par.}
 & SP=2 (F) & 200.4 (+18.8) & 200.3 (-5.7) & 217.1 (+2.7) \\
 & SP=4 (F) & 235.3 (+39.5) & 212.9 (+0.2) & 209.6 (-0.8) \\
 & SP=2 (S) & 198.7 (+17.8) & 212.7 (+0.1) & 217.0 (+2.7) \\
 & SP=4 (S) & 223.3 (+32.4) & 228.8 (+7.7) & 222.4 (+5.3) \\
\hline
\end{tabular}
\end{table}


\section{Discussion and Outcomes}
The purpose of this section is to translate experimental observations into practical guidance for practitioners training multimodal models on tightly coupled heterogeneous systems.

{\bf Optimizer offloading improves energy with overlap.}
Optimizer offloading consistently lowers total energy consumption compared to the no-offloading baseline across all model scales. On GH200, offloading serves not only as a memory-capacity solution but also, when paired with a high-bandwidth CPU-GPU interconnect, enhances efficiency by overlapping data movement with computation. This overlap directly improves energy by reducing idle periods. Offloading does not reduce instantaneous power; instead, it often increases GPU and module power by minimizing stalls and increasing overall system activity. 

{\bf Asynchronous execution is consistently more effective than synchronous execution.}
Among the optimizer and activation offloading strategies, asynchronous execution offers the most favorable balance among power stability, throughput, wall time, and energy consumption. The normalized step profile indicates that asynchronous optimizer offloading yields a smoother node-level power trace, particularly during the backward pass, which accounts for the majority of the computation. This smoother profile demonstrates improved overlap between communication and computation and reduced synchronization-induced stalls. These effects translate into higher energy efficiency by minimizing stall-induced energy waste. Similar patterns are observed in end-to-end metrics, where asynchronous runs consistently match or surpass synchronous runs in throughput and achieve the greatest energy savings. These results establish asynchronous execution as the preferred default runtime among the evaluated approaches.

{\bf Activation offloading becomes more valuable at larger scales.}
Activation offloading yields smaller improvements for the 7B model and larger improvements for the 32B and 72B models, suggesting that it is more effective as memory pressure increases. 
Another finding is that activation offloading affects energy consumption much more than it affects the immediate power draw. GPU average power remains fairly steady across different activation-offloading setups. In contrast, changes at the module level are more pronounced because the CPU, memory system, and interconnect are more involved in moving activations. Thus, energy efficiency gains arise primarily from reducing inefficient execution phases rather than lowering peak power. This means the main advantage of activation offloading is not reducing peak device power, but making inefficient execution periods shorter or smoother, which helps lower energy use. Activation offloading should be viewed as a way to save energy and time in memory-bound scenarios, especially for larger models, rather than as a power-capping method.

{\bf Sequence parallelism is highly effective for reducing time, but its efficiency depends on scale.}
Sequence parallelism can substantially reduce wall time, particularly when combined with GPU scaling. However, this comes at a communication cost, measured in TFLOP/s and in module-level power consumption. Energy use typically increases with higher SP degrees, especially for larger models, but the benefits level off at higher partitioning due to increased communication and synchronization overhead. As a result, energy depends on balancing reduced runtime against higher communication cost, making sequence parallelism a strong-scaling tool with trade-offs. It accelerates execution and often reduces total energy consumption, but at the cost of per-device efficiency. The effect depends on model size. For 7B, moderate SP is most effective, whereas excessive partitioning incurs overhead; for 32B and 72B, higher SP is beneficial, as it mitigates sequence and memory costs despite increased communication. So, the best SP setting depends on the workload. It should be adjusted based on model size, GPU count, and the communication speed. On GH200, the fast interconnect enables this trade-off.

{\bf GPU scaling alone is not an energy-saving solution.}
Using more GPUs can shorten wall time, but our experiments show this does not always lead to lower energy use. In the optimizer-offloading, scaling up without offloading does not match the energy savings achieved with offloading, even with greater parallelism. In sequence-parallel runs, scaling cuts execution time, but the extra communication can lower per-device efficiency and limit energy savings. Minimizing time-to-solution does not guarantee energy efficiency; on GH200, the best results are achieved with moderate scaling, yielding runtimes that reduce idle time and memory traffic.

Based on the above analysis, we summarize the following guidelines for energy-efficient multimodal training on tightly coupled CPU–GPU systems:

\begin{itemize}[leftmargin=4mm]

\item Energy efficiency is primarily determined by data movement and system interaction rather than raw compute performance. Optimizing communication and memory behavior is therefore more critical than maximizing FLOP/s.

\item Asynchronous optimizer and activation offloading should be preferred on high-bandwidth systems, as overlapping computation with data transfers reduces idle time and improves both energy-to-solution and throughput.

\item Offloading should be viewed as a latency-hiding mechanism rather than a way to reduce instantaneous power, as its benefits come from reducing time and minimizing stalls.

\item The effectiveness of offloading depends on model scale: it improves GPU utilization for smaller models and alleviates memory pressure for larger models by leveraging CPU memory and interconnects.

\item Sequence parallelism should be applied conservatively, as moderate levels improve scalability, while aggressive partitioning introduces communication overhead that can degrade both performance and energy efficiency.

\item Increasing the number of GPUs reduces runtime but does not necessarily improve energy efficiency, since power consumption and communication overhead can offset performance gains.

\item Data movement across memory hierarchies, including HBM, LPDDR, and CPU-GPU interconnects, significantly contributes to overall energy consumption. Therefore, optimizing data locality and minimizing data transfers are critical objectives.

\item The most efficient configurations balance computation, communication, and memory usage rather than optimizing a single metric such as throughput.

\item Achieving energy efficiency on GH200 requires hardware-aware scheduling that exploits CPU–GPU coupling and communication-aware execution.

\end{itemize}

\section{Conclusion}

This study shows that energy behavior in GH200 multimodal training is driven more by data movement across the memory hierarchy than by computation. While GPU instantaneous power dominates, energy differences arise from memory traffic, CPU–GPU coordination, and interconnect usage, especially at scale. Thus, optimizing for throughput alone is insufficient, and cross-layer, system-aware strategies are needed. Asynchronous offloading and tuned sequence parallelism improve performance and energy efficiency by overlapping computation and communication, but their benefits are scale-dependent and require balanced computation, communication, and memory usage. These results highlight that runtime and system co-design are essential, with efficiency driven by reducing idle time and data movement.

\section*{Acknowledgment}
This work was supported by King Abdullah University of Science and Technology (KAUST). We acknowledge the KAUST Supercomputing Laboratory (KSL) for providing access to the Ibex cluster for initial runs and GPU experiments. This project was granted access to the JUPITER supercomputer through the JUPITER Research and Early Access Program (JUREAP). JUPITER is funded by the EuroHPC Joint Undertaking, the German Federal Ministry of Research, Technology, and Space, and the Ministry of Culture and Science of North Rhine-Westphalia.


\bibliographystyle{IEEEtran}
\bibliography{bibliography}


\end{document}